\documentclass[twocolumn,amsmath,amssymb,superscriptaddress]{revtex4-1}
\usepackage{graphicx}
\usepackage{color}

\begin{document}
\title{Perspective: Nonequilibrium glassy dynamics in dense systems of active particles}

\author{Ludovic Berthier}

\affiliation{Laboratoire Charles Coulomb, UMR 5221 CNRS,
Universit{\'e} Montpellier, Montpellier, France}

\author{Elijah Flenner}
\affiliation{Department of Chemistry, Colorado
State University, Fort Collins, CO 80523, USA}

\author{Grzegorz Szamel}
\affiliation{Department of Chemistry, Colorado State University, 
Fort Collins, CO 80523, USA}

\date{\today}

\begin{abstract}
Despite the diversity of materials designated as active matter, virtually all active systems undergo a form of dynamic arrest when crowding and activity compete, reminiscent of the dynamic arrest observed in colloidal and molecular fluids undergoing a glass transition. We present a short perspective on recent and ongoing efforts to understand how activity competes with other physical interactions in dense systems. We first review recent experimental work on active materials that uncovered both classic signatures of glassy dynamics and intriguing novel phenomena at large density. We introduce a minimal model of self-propelled particles where the competition between interparticle interactions, crowding, and self-propulsion can be studied in great detail. We discuss more complex models that include some additional, material-specific ingredients. We end with some general perspectives on dense active materials, suggesting directions for future research, in particular for theoretical work.
\end{abstract}

\maketitle

\section{Introduction}

\subsection{Fluid-to-solid transitions}

The topic of this perspective is a widely observed phenomenon. Take a dense system of `particles', which can be molecules, droplets, cells, grains, or animals. When the density is not too large, these particles can easily move, and they can be fueled by thermal fluctuations, chemical reactions, internal motors, or muscles. The system is in a fluid-like state. As the density increases, it becomes increasingly difficult for the particles to find pathways that allow them to move over large distances. The competition between particle crowding in a dense environment and the energy injected at the particle scale may result in a transition from a fluid regime to a dynamically arrested regime where individual particles are permanently trapped by their neighbors. In this arrested state, the particles respond as a homogeneous block to external perturbations; the system has become a solid. Very simple systems, such as assemblies of identical particles in thermal equilibrium would easily crystallize at large densities, but for many `complex' particles the arrested state is fully  disordered. The phase transformation between an equilibrium fluid and an arrested amorphous state is the glass transition~\cite{today}. 

We argue below that the transition from a fluid to an amorphous solid is ubiquitously observed not only for molecules and small colloids (which form molecular and colloidal glasses), but is similarly relevant to describe a large class of active materials~\cite{review1,review2,review3}, where the `particles' can be phoretic colloids, self-propelled grains, or crawling cells. In those examples, that we review below, the competition arises between the crowding of the active particles (that tends to arrest them) and the intensity of the active forces (that make them move). 
 
In recent years, we noticed that the fluid-solid transition in active materials has often been described as a jamming transition~\cite{Mongera2018}, rather than a glass transition. We see two reasons for this. First, the word jamming itself is perhaps more easily grasped. Second, it echoes work performed in the granular matter community about 20 years ago that attempted to unify the physics of seemingly disparate physical systems, from molecules to grains and foams~\cite{jamming}. In a sense, cells, robots and phoretic colloids would only be more examples of the same type of physics. More recently, however, the distinction between the glass and the jamming transition, and the specific features associated to both phenomena have been clarified and explained in great detail~\cite{francesco}. Broadly speaking the competition between crowding and agitation is captured by the glass transition phenomenon. In contrast, jamming is understood as a purely geometric transition between viscous and rigid behavior in the absence of any kind of dynamics. Thus, jamming is a zero-temperature, or, for the purpose of the present paper, a zero-activity limit. Strictly speaking, therefore, active particles can undergo jamming only when they are not active.

\subsection{The equilibrium glass transition}

Let us first quickly review the main phenomenology associated to the equilibrium glass transition. 

The most noticeable phenomenon accompanying the incipient glass transition is the enormous slow down of the microscopic dynamics~\cite{BerthierBiroliRMP}. For instance, the viscosity of a hard sphere colloidal glass former can 
increase by seven orders of magnitude when the colloidal suspension's volume fraction changes from a dilute value of a few percent to a value close to the so-called colloidal glass transition~\cite{Cheng2002,Russel2013}. 
Even more impressively, a viscosity of a good molecular glass former can increase by twelve orders of magnitude upon decreasing the temperature by a mere factor of two~\cite{Angell1995}. 

This dramatic slowing down is only one of many spectacular changes in the dynamics which occur when the glass transition, either in a colloidal or a molecular liquid, is approached. For example, the very nature of the single-particle 
motion changes~\cite{RouxHansen,LoewenHansen}. Whereas the single particle motion in a low to moderately high volume fraction suspension can 
be well described by a diffusion process, at low temperature any given particle goes through a series 
of different dynamics. First, before it becomes aware of its surroundings, it freely explores its immediate 
neighborhood. Next, on intermediate time scales, it becomes caged by its solvation shell. Finally, on
much longer time scales, it manages to escape from the cage. After many such transient  
localizations, the long-time motion may be described by an effective diffusion process, albeit with 
a much smaller diffusion coefficient. 

This two-step single particle motion affects the behavior of all time correlation functions, which also exhibit two-step decay with an intermediate time plateau reflecting the cage dynamics of the particles~\cite{RouxHansen,LoewenHansen}. This microscopic dynamics suggests that fluids approaching the glass transition display viscoelastic response to an applied stress.  

In addition to the caged and slow dynamics, a large combination of ingenious experiments~\cite{EdigerAnnRev}, computer simulations~\cite{Kobdynhet,Lacevic2003} and theoretical analyses~\cite{BerthieretalJCP}, has established that glassy dynamics is also increasingly heterogeneous~\cite{Cipellettibook}. Dynamic heterogeneity means that in a viscous liquid 
there is a coexistence of fast particles, with a motion that is much faster than the average, and slow particles, with a motion that is much slower than the average. However, over timescales that are (commonly but not universally) considered much longer than the typical relaxation time of the fluid, fast and slow populations lose their character and dynamic exchanges between the two sets of particles can occur. The additional and, in fact, crucial aspect of dynamic heterogeneity is that fast and slow particles are also spatially correlated over a new, dynamic correlation length scale that grows upon approaching the glass transition. 

Physically, dynamic heterogeneity can be understood as a direct consequence of particle crowding. In order to perform some motion, a particle closely surrounded by its neighbor needs to coordinate motion with its neighbor in order to diffuse. This is very natural. After all, human beings in a dense crowd spontaneously coordinate their motion to become mobile.  
Collective motion is an important aspect of the physics of active materials. It is therefore important to realize that particle crowding is a key ingredient to trigger correlated motion, even for systems at thermal equilibrium.

\subsection{Glassy dynamics in active matter}

In the last decade it has been realized that many, if not all, of the phenomena associated with glassy dynamics
could also be observed in dense active matter systems. For the purpose of this article, the term 
active matter encompasses a variety of different materials~\cite{review1,review2,review3}. They range from living tissues to systems of active 
colloidal particles to macroscopic granular objects driven by mechanical perturbations. The differences 
between these very diverse systems have consequences for the phenomena that can be observed and for the
details of the corresponding experiments. 

Let us start with some specific experiments on cells and tissues. Typically, in these systems cells are proliferating and sometimes
also dying, with the overall cell density being a non-trivial function of time during the experiment. Since 
the dynamics of dense systems is very sensitive to their density, the fact that the number density is changing
imposes additional variation upon experimental results. 

Angelini \textit{et al.} \cite{Angelini2011} studied the dynamics of a confluent epithelial cell sheet. They monitored 
cell motion over a broad range of length scales, time scales and cell densities. They found that 
with increasing cell density, the dynamics slow down. The log of the inverse self-diffusion coefficient
was found to have non-linear dependence on the cell density, which shares some vague analogy with the non-linear 
dependence of the log of the relaxation time on the inverse temperature. 
Even more interestingly, Angelini \textit{et al.} found
that not only the dynamics were slowing down but also they were increasingly more heterogeneous. They
estimated the dynamic correlation length and found that it increased with increasing cell density.

Garcia \textit{et al.} \cite{Garcia2015} studied a different confluent epithelial cell sheet. They also found a slowing
down of the dynamics upon increasing the cell density. They investigated dynamic heterogeneity and
determined the dynamic correlation length. Interestingly, upon increasing the cell density (which increased
during the duration of the experiment) the length exhibited a non-monotonic behavior, first increasing
and then decreasing with increasing density. Garcia \textit{et al.} also found a distinctly non-equilibrium
feature of active glassy dynamics; non-trivial equal time velocity correlations. We recall that in equilibrium
systems, either colloidal or molecular, equal-time velocity correlations are trivial; velocities of different
particles are uncorrelated. 
In contrast, Garcia \textit{et al.} 
determined that the length scale  characterizing these correlations also 
exhibits a non-monotonic dependence
on the cell density. Notably, the dynamic correlation length and the velocity correlation length were found to 
be correlated; their relation was found to be monotonic, in spite of the complex dependence of each length itself on time. 

More recently, Mongera \textit{et al.} \cite{Mongera2018} performed a more complex series of experiments. 
They focused on the important biological process of vertebrate body axis elongation. They identified and investigated 
an amorphous solidification process in which the cells become solid-like as they transition from mesodermal 
progenitor zone (MPZ) to presomitic mesoderm (PSM). The transition was monitored through 
the mean-square displacement, which was found to increase in a diffusive way in the the MPZ and
exhibit arrest in the PSM. They also studied mechanical response of both PSM and MPZ and identified
a yield stress, which is a commonly observed mechanical property of amorphous solids. Finally, they 
investigated active fluctuations and the role they play in the amorphous solidification. They found
that active fluctuations are strong in the MPZ and weak in the PSM, in analogy with thermal 
fluctuations in liquid and glassy phases. This finding led them to hypothesize that these active 
fluctuations play the role of a temperature. As we mentioned earlier, in our view the amorphous solidification
process uncovered by Mongera \textit{et al.} is a glass rather than a jamming transition, since it happens at
a non-vanishing level of the activity.

Colloidal systems have long served as a laboratory to observe and understand glassy 
dynamics~\cite{HunterWeeks,GokhaleAinP}. The reason is that many
colloidal experiments allowed workers to obtain significantly more information about the microscopic dynamics of
the colloidal systems than in atomic systems. The wealth of available information more than compensates for
the fact that in colloidal systems the slowing down is not as spectacular as in atomic systems and typically
only five or six decades of the change of the relaxation times can be observed. Colloidal systems consisting
of active Janus colloids (in which one part of the colloidal particle is covered with some kind of catalyst,
leading to a self-propelled motion) were one of the first synthetic active matter systems. Initially, the 
experiments focused on single-particle motion and then on moderately dense systems~\cite{Howse,Palacci,Buttinoni}. In the latter systems, 
clustering and phase separation of active colloidal particles with purely repulsive interactions can be 
observed~\cite{Buttinoni,Theurkauff}. More recently, some groups started investigating the structure and dynamics of dense active
colloidal systems~\cite{Klongvessa2019}. Although the details of the experiments are just emerging, it is clear that the dynamics of 
dense active colloidal systems exhibit classic signatures of glassy colloidal dynamics, with non-trivial dependencies of the microscopic 
dynamics upon changes in control parameters. We hope that further
work in this area will yield a wealth of information on the interplay between colloidal crowding and phoretic activity, since such experiments probe an simpler version of the more complex tissue dynamics. 

Finally, we mention a new experimental active matter system that belongs to the category of driven granular 
systems. For some time Dauchot's group has used macroscopic grains driven by shaking the plate on which they are 
placed as an experimental model system to study glassy dynamics and jamming phenomena~\cite{CoulaisSM}. Strictly speaking, this
driven system is active in the sense that it is an athermal system, devoid of any intrinsic dynamics and 
driven at the level of individual particles. However, since the drive is memory-less and the grains and the
motion are isotropic, the most appropriate theoretical and/or simulational model for this system is an
effective equilibrium system with thermal fluctuations. 
Recently, Dauchot's group introduced two new model systems of shaken grains. The first system consist of
monodisperse polar grains whose asymmetry leads to persistent motion under shaking~\cite{Deseigne}. These grains, therefore,
behave very much as self-propelled particles with ballistic short-time motion and effectively diffusive long-time motion. 
The system of polar grains is best modeled by models used for polar active matter~\cite{model-olivier,model-olivier2}.
More recently, a bidisperse version of the polar grain system was also introduced. In this system, crystallization is suppressed, and active glassy dynamics can be observed. 
Preliminary results again suggest important slowing down of the final diffusive motion accompanied
by an intermediate-time localization of individual particles. We hope that further analysis will investigate the presence of correlated motion or velocity correlations in this system.


In all these active systems, particle motion is uniquely controlled by the intensity of the active forces of biological, chemical, or mechanical origin. The particles are fully arrested when activity is absent and may start to diffuse and undergo interesting dynamics for a finite level of activity. The main questions we wish to address are as follows. How should we describe the transition between a dense amorphous solid and a fluid controlled by active forces? What is the microscopic dynamics that can be expected at the transition? Can this transition be analyzed theoretically using simplified models of active matter?

This brief review is focused on models of active glassy dynamics. We start by a short review 
of the phenomena observed in equilibrium glassy systems. Next, we discuss the minimal
ingredients of models of active glass-forming systems and introduce what we consider the minimal model
that can exhibit the salient features of active glassy dynamics. We then show some of the unexpected phenomena 
observed in simulations and (sometimes) deduced from theoretical
considerations. We also briefly discuss models that attempt to model specific experimental systems more 
closely, albeit at the cost of introducing more control parameters and more variables. We end with some perspectives for future research in the field.

\section{Glassy dynamics in non-active systems}

\subsection{Short review of the equilibrium glass transition}

To set the stage for the discussion of active glassy dynamics, we first discuss the salient features of 
the structure and dynamics of equilibrium (non-active, \textit{i.e.} `passive') glassy systems. We  
note that many of the microscopic phenomena discussed in this section require detailed information about 
particles' motion on the microscopic scale, and for that reason they were first observed in computer
simulations and later studied in colloidal systems. We also recall that almost all good glass-formers
studied in computer simulations are many-component mixtures (for single component systems with typical
interaction potentials it is virtually impossible to avoid crystallization upon even mild supercooling). 

All the examples shown in this section and in the next one were obtained for a 50:50 binary mixture of 
spherically symmetric particles interacting via the Lennard-Jones potential cut at the minimum, which is usually
referred to as the Weeks-Chandler-Andersen (WCA) interaction~\cite{WCA},
\begin{equation}
\label{potential}
V_{\alpha \beta}(r) = 4 \epsilon
\left[ \left( \frac{\sigma_{\alpha \beta}}{r} \right)^{12}
- \left( \frac{\sigma_{\alpha \beta}}{r} \right)^6 \right],
\end{equation}
for $r \le \varsigma_{\alpha \beta} = 2^{1/6} \sigma_{\alpha \beta}$
and zero otherwise. In Eq.~\eqref{potential}, $\alpha, \beta$ denote the particle species
$A$ or $B$, $\epsilon =1$ (which sets the unit of energy),
$\sigma_{AA} = 1.4$, $\sigma_{AB} = 1.2$, and
$\sigma_{BB} = 1.0$ (which sets the unit of length). In all the figures except for Figs. \ref{fig:pdistrdispl} and \ref{fig:noneq}c we show properties pertaining to
the larger, $A$, particles. In Figs. \ref{fig:pdistrdispl} and \ref{fig:noneq}c we show probability distributions pertaining to the smaller, $B$, particles.

A defining feature of an approaching glass transition is a dramatic slowing down of a liquids dynamics with 
little change of the pair structure upon a small change of the temperature $T$ and/or density. The pair 
structure is typically monitored through the pair correlation function~\cite{HansenMcDonald}
\begin{equation}
\label{eq:pair}
g(r) = \frac{1}{\rho N} \left< \sum_{n,m\ne n} \delta[\mathbf{r} - ( \mathbf{r}_n(0) - \mathbf{r}_m(0)) ] \right>
\end{equation}
or the static structure factor~\cite{HansenMcDonald}
\begin{equation}
\label{eq:structure}
S(q) = \frac{1}{N} \left< \sum_{n,m} e^{i\mathbf{q} \cdot (\mathbf{r}_n(0) - \mathbf{r}_m(0))} \right>.
\end{equation}
In Eqs.~(\ref{eq:pair}) and (\ref{eq:structure}),
$N$ is the number of particles, $\rho$ is the number density, and $\mathbf{r}_n(t)$ is the 
position of particle $n$ at a time $t$. While these functions are related through a Fourier transform 
and thus encode the same information,  
it is easier to distinguish differences in structure on nearest neighbor length scales by examining 
$g(r)$ and it is easier to compare the decay of the structure on longer length scales by examining $S(q)$. 

\begin{figure}
\includegraphics[scale=0.3]{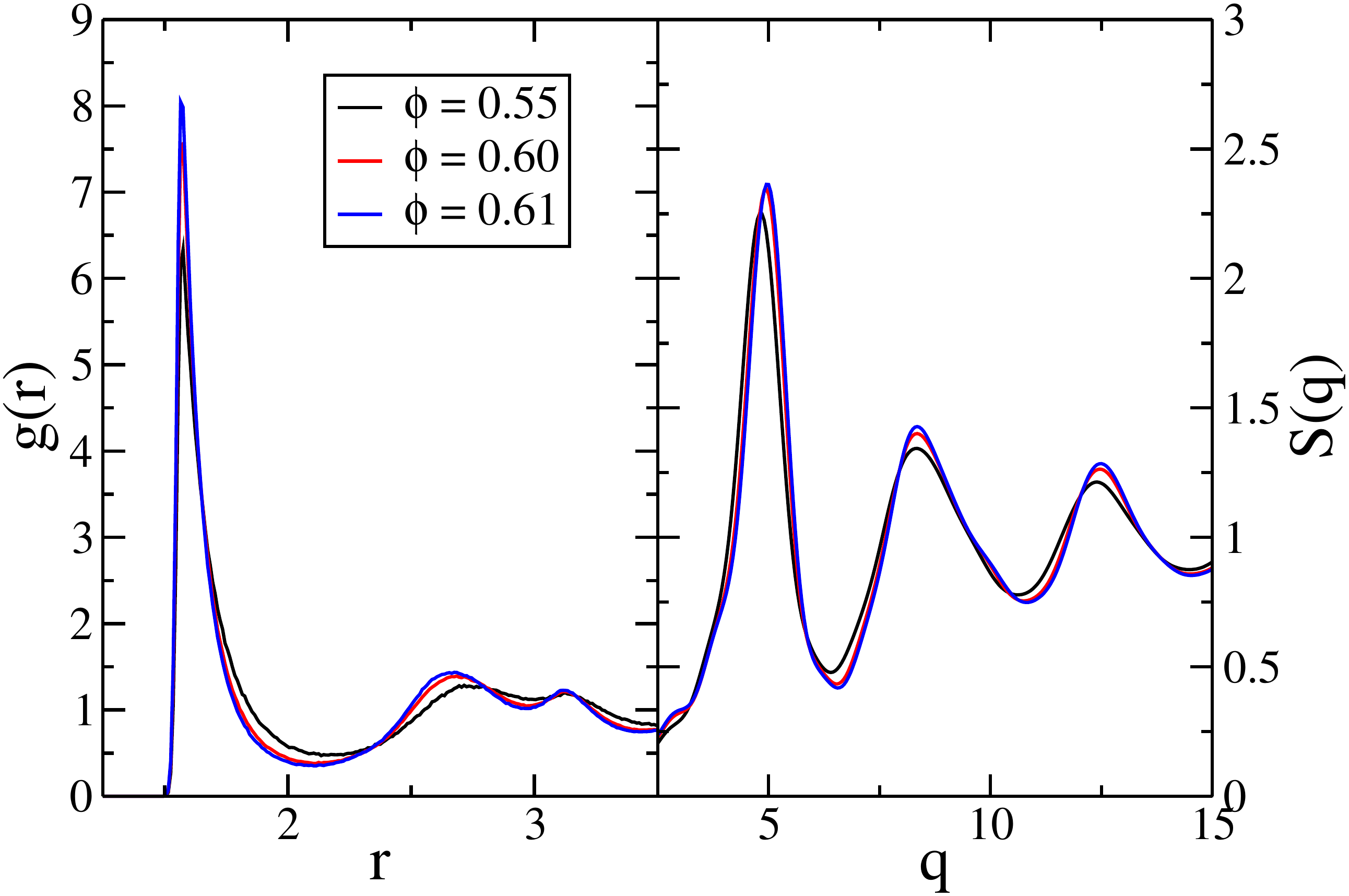}
\caption{\label{fig:pstatics} The pair correlation function, $g(r)$, (left panel) and the static
structure factor, $S(q)$, (right panel), for three volume fractions in the vicinity of the glass transition for 
an equilibrium WCA system at a low temperature, $T=0.01$.}
\end{figure}

In Fig.~\ref{fig:pstatics} we show the density dependence of the pair correlation function and the static structure
factor of a WCA system at the constant temperature, $T=0.01$. While the
pair structure changes very little, the long-time dynamics of the system (as characterized by time correlations defined below) 
slows down by approximately 3 orders of magnitude. 

To determine if the structure evolves in time we can examine time dependent versions of Eqs.~(\ref{eq:pair}, \ref{eq:structure}) where $\mathbf{r}_n(0)$ is replaced by $\mathbf{r}_n(t)$. 
The time dependent version of Eq.~(\ref{eq:structure}) defines the collective (coherent) intermediate scattering 
function~\cite{HansenMcDonald}
\begin{equation}
\label{eq:collective}
F(q;t) = \frac{1}{N} \left< \sum_{n,m} e^{i\mathbf{q} \cdot [\mathbf{r}_n(t) - \mathbf{r}_m(0)]} \right>,
\end{equation}
which characterizes the relaxation of the initial structure on a length scale characterized by $q = |\mathbf{q}|$.
The characteristic decay time
of $F(q;t)$ for wavevector $q$ near the peak position of the static structure factor defines a structural relaxation time, usually referred to as the $\alpha$ relaxation time, $\tau_\alpha$.
For reasons of computational efficiency, quite often one monitors the self-intermediate scattering function~\cite{HansenMcDonald} 
\begin{equation}
\label{eq:self}
F_s(q;t) = \frac{1}{N} \left< \sum_{n} e^{i\mathbf{q} \cdot [\mathbf{r}_n(t) - \mathbf{r}_n(0)]} \right>,
\end{equation}
which corresponds to the $n=m$ terms in Eq.~(\ref{eq:collective}). The characteristic decay time
of $F_s(q;t)$ for $q$ at the peak position of the static structure factor is nearly equal to $\tau_\alpha$, but easier to compute.

\begin{figure}
\includegraphics[scale=0.3]{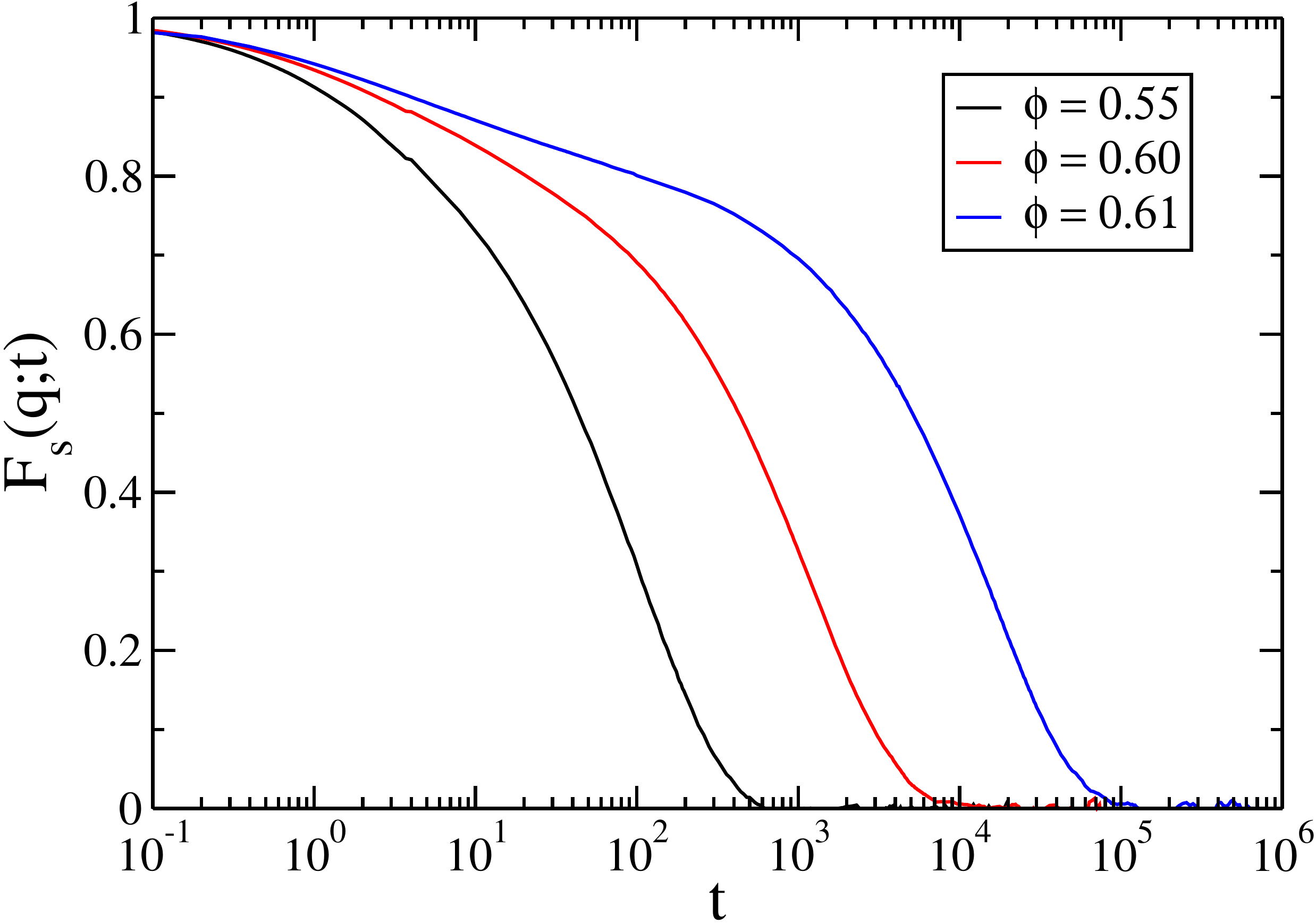}
\caption{\label{fig:pdynamics} The self-intermediate scattering function, $F_s(q;t)$, 
for three volume fractions in the vicinity of the 
apparent glass transition, for an equilibrium WCA system at a low temperature, $T=0.01$. The 
wavector $q$ is close to the position of the first peak of the static structure factor, $q=5$.
Small changes of the static structure shown in Fig. \ref{fig:pstatics} are concurrent with a 
dramatic slowing down of the dynamics.}
\end{figure}

In a simple liquid above the onset of glassy dynamics, it is found that $F_s(q;t)$ 
decays nearly exponentially. This is consistent with
a Gaussian distribution of particle displacements
\begin{equation}
G_s(r;t) = \frac{1}{\rho N} \left< \sum_{n} \delta[\mathbf{r} - ( \mathbf{r}_n(t) - \mathbf{r}_n(0)) ] \right>
\end{equation}
whose mean-square average increases linearly in time, \textit{i.e.} with Fickian diffusion.

Two differences occur when the liquid is supercooled or if its density is increased.  
A plateau develops in $F_s(q;t)$ that indicates that the particles are localized, as a solid, 
at intermediate time scales. The particles are said to be trapped in a cage formed by their neighbors, and
they have to escape this cage for $F_s(q;t)$ to decay from the plateau.
The decay from the plateau occurs at increasingly later times upon approaching the glass transition, and 
an operational definition of the glass transition is that one is no longer willing to wait for $F_s(q;t)$ to 
decay from this plateau. The other major change is that the decay after the plateau
is no longer exponential; it is usually fitted by a stretched exponential function, $\propto \exp(-(t/\tau)^\beta)$,
where the so-called stretching exponent $\beta$ decreases with decreasing temperature. 
In Fig. 2 we show $F_s(q;t)$ for the same state points as in Fig.~\ref{fig:pstatics}. 

\begin{figure}
\includegraphics[scale=0.3]{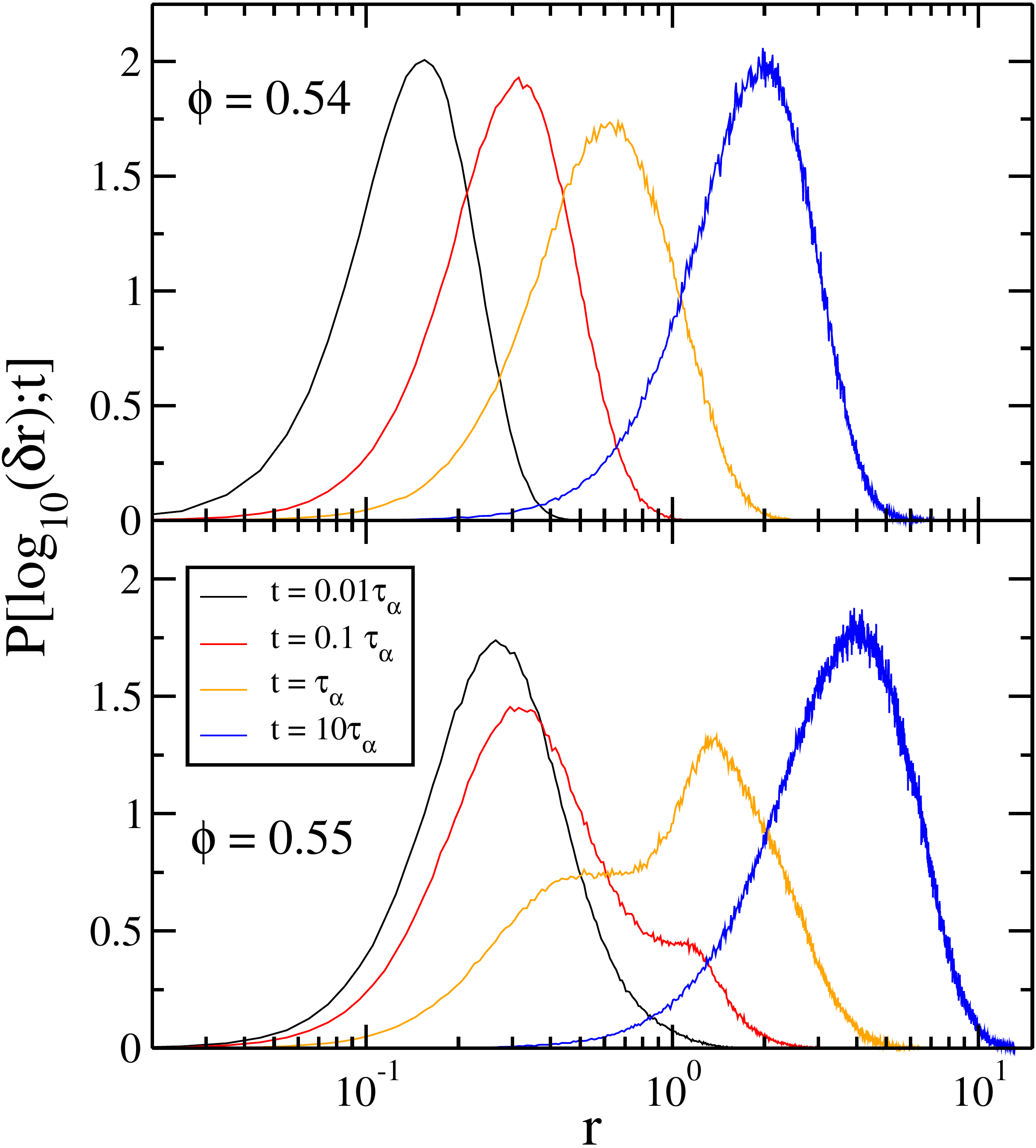}
\caption{\label{fig:pdistrdispl} The probability of the logarithm of the single-particle displacements,
$P[\log_{10}(\delta r);t]$
for two volume fractions in the vicinity of the 
apparent glass transition, for an equilibrium WCA system at a low temperature, $T=0.01$. For each volume fraction 
$P[\log_{10}(\delta r);t]$ is shown at times $t$ equal to $0.01$, $0.1$, $1$ and $10$ $\alpha$ relaxation times 
$\tau_\alpha$ at that volume fraction. 
For Gaussian distributions of displacements the shape of $P[\log_{10}(\delta r);t]$ does 
not depend on time and its peak value equals to $2.13$.
Small changes of the static structure shown in Fig. \ref{fig:pstatics} are concurrent with
dramatic changes of the dynamics.}
\end{figure}

The non-exponential decay of $F_s(q;t)$ implies that the probability of the displacements $G_s(r;t)$ is non-Gaussian. 
The non-Gaussian character of the single particle displacements was investigated in some detail. 
It was found that the particles are localized 
for an extended period of time, then make a relatively quick jump to another cage where they stay
for another extended period of time. A consequence of this hopping-like motion
is that the particles can be separated into slow and fast sub-populations. The slow particles are ones that moved less than
expected for a Gaussian distribution of displacements and the fast particles are ones that
moved more than what was expected for a Gaussian distribution of displacements. Importantly,
the slow and fast particles are also found to be spatially correlated and form increasing larger
clusters upon approaching the glass transition. These spatially heterogeneous dynamics 
is recognized as one of the hallmarks of glassy dynamics. 
To inspect the Gaussian character of the single-particle motion it is convenient 
to monitor the probability distribution 
of the logarithm of single-particle displacements, $\log_{10}(\delta r)$,
during time $t$, $P[\log_{10}(\delta r);t]$, which is simply related to $G_s(r;t)$,
$P[\log_{10}(\delta r);t] = \ln(10) 4 \pi \delta r^3 G_s(\delta r,t)$ \cite{Puertas,Flenner2005a}. 
The usefulness of $P[\log_{10}(\delta r);t]$ comes from the fact that if $G_s(r;t)$ is Gaussian,
then the shape of the probability distribution $P[\log_{10}(\delta r);t]$
is independent of time and its peak is equal to
$\ln(10) \sqrt{54/ \pi}\;e^{-3/2} \approx 2.13$.
In Fig.~\ref{fig:pdistrdispl} we show $P[\log_{10}(\delta r);t]$ observed for a simple fluid, which indicates
near-Gaussian diffusion (top), and the bimodal $P[\log_{10}(\delta r);t]$ distribution obtained at 
intermediate timescales for a slowly relaxing system (bottom).  

In the description above, we have not specified the type of microscopic dynamics giving rise to the glassy dynamics. Interestingly, it was demonstrated by direct numerical comparison that the global evolution of the relaxation dynamics, of the slow relaxation of time correlation functions, of the dynamic heterogeneity associated to spatio-temporal fluctuations of the dynamics are actually the same for Newtonian, Brownian, Langevin, or even Monte Carlo dynamics. Physically, this implies that details of the microscopic motion at very short times do not affect the manner in which the slow dynamics proceeds at much larger times. In other words, the strong separation of timescales makes the details of the driving dynamics irrelevant at long times. This finding will play an important role when discussing the role of non-equilibrium active forces.

\subsection{Driven dynamics of glasses: Rheology}

\label{sec:rheology}

As we wish to understand the behavior of dense materials driven by active forces, it is interesting to mention that glassy materials can be driven out of equilibrium by many types of forces, and `active' forces are only one particular example on which we shall focus below.

A well-known example of a driving force that is frequently applied to a dense assembly of particles is an external mechanical perturbation that can take the form of a shear flow, or a constant stress~\cite{Larson}. The obvious qualitative difference with active forces is that such mechanical perturbation is applied at large scale, rather than at the particle level.    
Before discussing the later, it is therefore interesting to learn from the former case. 

The field of glassy materials driven by an external mechanical constraint relates to the rheology of glassy systems. Starting from an arrested glass at low temperature, the application of a constant force (such as a shear stress) may give rise to a yielding transition. Whereas the glass responds in a nearly linear manner at small applied force, the response becomes non-linear at larger applied force until a well-defined force threshold is crossed (called a ``yield stress'') above which the glass deforms plastically and undergoes microscopic relaxation. The yielding is thus a form of a solid-to-fluid transition driven by an external force of sufficient strength, which is currently under intense scrutiny~\cite{Ozawa,yieldRMP}. The response of a glassy system to an applied force is obviously relevant to active glassy materials. 

Another way to mechanically drive a glass is to impose a constant rate of deformation, \textit{i.e.} a finite shear rate. This is again a useful analogy since such geometry introduces a new timescale (the shear rate) for the external driving force, in close analogy with self-propelled motion in active particle systems. 
The presence of a finite shear rate has been analyzed in great detail~\cite{BB,MRY}. The main finding is that to sustain a constant deformation rate, a dense system needs to constantly undergo plastic rearrangements, and the structure is thus never dynamically arrested. In other words, the system is always in a driven steady state where particles diffuse and the structure rearranges and there cannot be a fluid-to-solid transition since the material is permanently in a non-equilibrium fluid phase. 
 
\subsection{Fluid-to-solid jamming transition at zero temperature}

\label{sec:jamming}

As briefly mentioned in the introduction, the jamming transition describes a fluid-to-solid transition in the absence of any fluctuations, in particular thermal fluctuations~\cite{LiuNagelreview}. 

A clean setting to observe the jamming transition is to consider packings of soft repulsive spheres; imagine for instance green peas. Peas are a useful image, as thermal fluctuations are clearly insufficient to drive their dynamics. The jamming transition separates a low-density regime where the assembly of peas can not sustain a shear stress and responds as a fluid, from a large-density regime where the assembly of peas responds as a solid. For repulsive spheres, the details of the jamming transition have been worked out in great detail. In particular, it is found that the emergence of rigidity corresponds to a nonequilibrium critical point, characterized by power laws and several critical exponents. In particular, the pair correlation function $g(r)$ in Eq.~(\ref{eq:pair}) develops singular behavior exactly at the jamming transition, the yield stress increases continuously from zero as a power law of the density, etc. 

\begin{figure}
\includegraphics[width=0.9 \columnwidth]{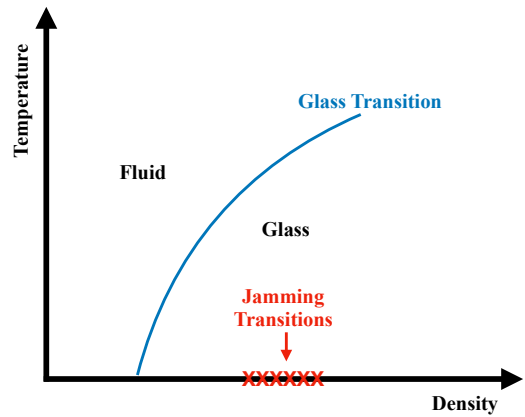}
\caption{\label{fig:rhot} 
Schematic temperature-density phase diagram for soft repulsive spheres undergoing a glass transition at finite temperature, and jamming transitions in the absence of thermal fluctuations.}
\end{figure}

Since the transition takes place at zero temperature, it is important to realize that there is, by definition, no glassy dynamics that can be observed near the jamming transition, since the former can only emerge when particles are driven by some sort of fluctuations. Sheared assembly of such non-Brownian particles does not display glassy dynamics either. 
Another important consequence of the absence of any dynamics is that the preparation protocol of the athermal packing needs to be specified to analyze jamming~\cite{ohern,torquato}. In particular, it is found that the location of the jamming transition for a given system cannot be unique, but is instead dependent on the packing preparation. There is thus not a unique jamming transition density, but instead a line of critical protocol-dependent jamming transitions~\cite{pinaki}. 

Although glass and jamming transitions both  describe fluid-to-solid transitions, they are quite distinct physical phenomena. This is most easily realized by studying the temperature-density phase diagram of soft repulsive spheres~\cite{tom,atsushi,hugo}, see Fig.~\ref{fig:rhot}. The jamming transitions take place at $T=0$ along the density axis over a range of densities. By contrast, the glass transition with its associated glassy dynamics takes place at finite temperature, and the structural and dynamical signatures associated to both transitions are observed in different, non-overlapping physical regimes. 

\section{Minimal model of glassy dynamics of active particles}

\label{minimal}

\subsection{Self-propelled particles}

To study how `activity' interferes with `crowding', our central theme, a minimal model should simultaneously capture the physics of crowding in dense particle assemblies, and those particles should be driven by active forces. 
Before modeling systems as complex as epithelial tissues or self-phoretic colloidal particles in a solvent, we suggest it is useful to learn some lessons from minimal models. In equilibrium, 
the glass transition is typically studied as a function of two control parameters, the particle density controlling crowding, and the temperature that drives the microscopic motion of the particles. 
Many investigations fix one of these control parameters 
and vary the other in order to simplify studying the phase diagram, and both directions are essentially equivalent.  

Active systems composed of self-propelled particles are characterized by two more control parameters, the persistence time of the active force and its average strength. Thus, the parameter space immediately doubles from two 
to four dimensions and the problem becomes intractable.
Since we are interested in how activity influences the glass transition, we can simplify matters by removing the effects of the thermal bath, \textit{i.e.} temperature, from the picture. Additionally we can either fix
the density or fix one of the parameters that controls the active motion to examine the influence of activity on the glass transition.

Self-propulsion in active matter model can take many forms, which are believed to yield to essentially the same behavior as far as collective behavior is concerned. A mathematically appealing minimal active matter model is a system of interacting active Ornstein-Ulenbeck particles (AOUPs) 
\cite{Szamel2014,Maggi,Fodor2016} where
particles perform overdamped motion in a viscous fluid, thus neglecting thermal fluctuations.
The self-propulsion forces evolve in time according to the 
Ornstein-Uhlenbeck~\cite{VanKampen} stochastic process. Thus, the equation of motion for the position $\mathbf{r}_n$ of particle $n$ is
\begin{equation}\label{eom1}
\dot{\mathbf{r}}_n = \xi_0^{-1}[\mathbf{F}_n + \mathbf{f}_n],
\end{equation}
where $\mathbf{F}_n = -\sum_{m\ne n} \nabla V(r_{nm})$ is the force
originating from pairwise particle interactions and $\mathbf{f}_n$ is the
self-propulsion force acting on particle $n$. Notice that since thermal fluctuations are neglected, there is no term 
corresponding to a thermal bath in Eq.~(\ref{eom1}), and thus  
without the active force the particles would only evolve towards the closest potential energy minimum. The pair potential $V(r)$ can be any simple model for a dense fluid usually studied in the field of simple glasses, from hard spheres to WCA and Lennard-Jones potentials. 

The equation of motion for the active force $\mathbf{f}_n$ is
\begin{equation}\label{eom2}
\tau_p\dot{\mathbf{f}}_n = -\mathbf{f}_n + \boldsymbol{\eta}_n,
\end{equation} 
where $\tau_p$ is the persistence time of the self-propulsion and $\boldsymbol{\eta}_n$
is an internal Gaussian noise with zero mean and variance 
$\left< \boldsymbol{\eta}_n \boldsymbol{\eta}_m \right>_{\mathrm{noise}} = 2 \xi_0 T_{\mathrm{eff}} \mathbf{I}\delta_{nm}\delta(t-t^\prime)$;
$\mathbf{I}$ denotes the unit tensor.
The average $\left< \ldots \right>_{\mathrm{noise}}$ denotes averaging over the noise distribution. 
The parameter $T_\mathrm{eff}$, which we will refer to as the (single-particle) effective temperature, 
quantifies the noise strength and, therefore, the magnitude of the 
active forces.

\subsection{Lessons from the dilute limit}

Before discussing dense systems it is useful to consider the dynamics of a single particle evolving
according to Eqs.~(\ref{eom1}-\ref{eom2})~\cite{Szamel2014}. The mean squared displacement of a single AOUP can be calculated as  
\begin{equation}
\left< \delta r^2(t) \right> 
= \frac{6 T_\mathrm{eff}}{\xi_0} \left[\tau_p \left(e^{-t/\tau_p} -1 \right) + t\right],
\end{equation}
which exhibits typical features of a persistent random walk. Indeed, 
for $t \ll \tau_p$ we can expand the exponential, $\left< \delta r^2(t) \right> 
\approx (3 T_\mathrm{eff} \tau_p/\xi_0) t^2$
and the motion is ballistic. For $t \gg \tau_p$ the exponential can be neglected, $\left< \delta r^2(t) \right> 
\approx (6 T_\mathrm{eff} /\xi_0) t$
and the motion is diffusive with a diffusion coefficient $D_0 = T_\mathrm{eff}/\xi_0$.
Here we see the origin of the name effective temperature: $T_\mathrm{eff}$ plays the same role as the
equilibrium temperature $T$ in the expression for the long-time diffusion coefficient of an isolated particle. 
Importantly, systems with the same effective temperature will have the same long time
diffusion coefficient in the absence of interactions. This makes $T_\mathrm{eff}$ a useful parameter to determine 
how the long-time dynamics changes upon approaching the glass transition. The persistence time $\tau_p$ 
gives the timescale for the transition from ballistic to diffusive motion for an isolated particle. 

After the introduction of the effective temperature $T_\mathrm{eff}$, it is natural to ask 
whether this parameter has other properties of the temperature  
in equilibrium passive systems. This question can be asked several ways. For example, one could ask
whether there is a linear response relation involving a single AOUP in which the equilibrium temperature 
$T$ is replaced by $T_{\mathrm{eff}}$? One could also ask whether 
the familiar Gibbs-Boltzmann distribution is recovered when a single AOUP is placed in an external potential, with the equilibrium temperature
replaced by $T_{\mathrm{eff}}$.

The answers to the above questions vary. It is possible to come up with a single
particle linear response problem in which, in the small frequency limit, the response and correlation
functions are related by $T_{\mathrm{eff}}$. Also, one can show that a single AOUP in a linear 
potential with a lower wall (the sedimentation problem), the probability
distribution has the Gibbs-Boltzmann form with the equilibrium temperature replaced by  $T_{\mathrm{eff}}$.
However, one can also show that the probability distribution of a single AOUP placed in a harmonic potential
has a Gaussian form, but the parameter that replaces the equilibrium temperature is in fact a function
of both $T_{\mathrm{eff}}$ and the persistence time $\tau_p$. These results suggest that, in general, $T_{\mathrm{eff}}$ does not always play the same role as the temperature in equilibrium systems. We note 
that in systems of interacting AOUPs, other temperature-like parameters could be defined~\cite{Levis,TeffEPL}. These temperatures will be influenced by the single particle effective temperature $T_{\mathrm{eff}}$, persistence time $\tau_p$
and the interparticle interactions. 


This, however,  does not preclude using the single particle effective temperature 
$T_{\mathrm{eff}}$ as a  control parameter for dense active suspensions since it can still tell us  
how much the interactions slow down in the long-time dynamics. Therefore, the minimal model of active glassy dynamics 
involves the single particle effective temperature $T_{\mathrm{eff}}$, the persistence time $\tau_p$ 
and the number density as control parameters. This set of control parameters allows us to investigate
the influence of \textit{being driven by non-equilibrium active forces} on the glassy dynamics.
In the limit of vanishing persistence time, the equations of motion (\ref{eom1}-\ref{eom2}) reduce to the equilibrium dynamics of an
overdamped Brownian system at the temperature equal to the effective temperature. Thus, the departure from equilibrium is quantified by the persistence time, and increasing the persistence time drives the system further away from equilibrium.
For the sake of brevity, we will sometimes use the phrases increasing/adding
activity to indicate increasing the persistence time.  Note, finally, that for the hard sphere interaction, the absolute value of $T_{\rm eff}$ does not compete with any energy scale, and the system is left with only two control parameters, density and persistence time. 

\subsection{Many-body physics at large density}

\subsubsection{Basic observations: Nonequilibrium glass transition}

Armed with a simple model of active particles, we can now examine if the glass
transition exists and how it evolves with changing $T_{\mathrm{eff}}$, 
the persistence time and the density. Initial studies of hard and soft spheres suggested
that adding activity does not destroy the glass transition, but rather pushes the 
transition to a higher density, in the case of hard spheres~\cite{Ni}, or to a lower temperature 
at constant density, in the case of soft spheres~\cite{Mandal2016}. 

It may appear logical that a driven system has a delayed glass transition, as compared to its equilibrium counterpart. We will show below a counterexample that proves that intuition incorrect. We recall that another incorrect intuition could be drawn from the analogy with driven glassy systems discussed in Sec.~\ref{sec:rheology} above, where we showed that a glass driven with a given deformation rate does not possess a glass transition and is always in a non-equilibrium steady state.
Simulations and theoretical analysis for self-propelled particles show that the local (as opposed to global mechanical deformation) nature of the driving in fact qualitatively changes the picture. A self-propelled particle system does undergo dynamic arrest to an amorphous glass that we call a {\it nonequilibrium glass transition}. 

\begin{figure}
\includegraphics[scale=0.3]{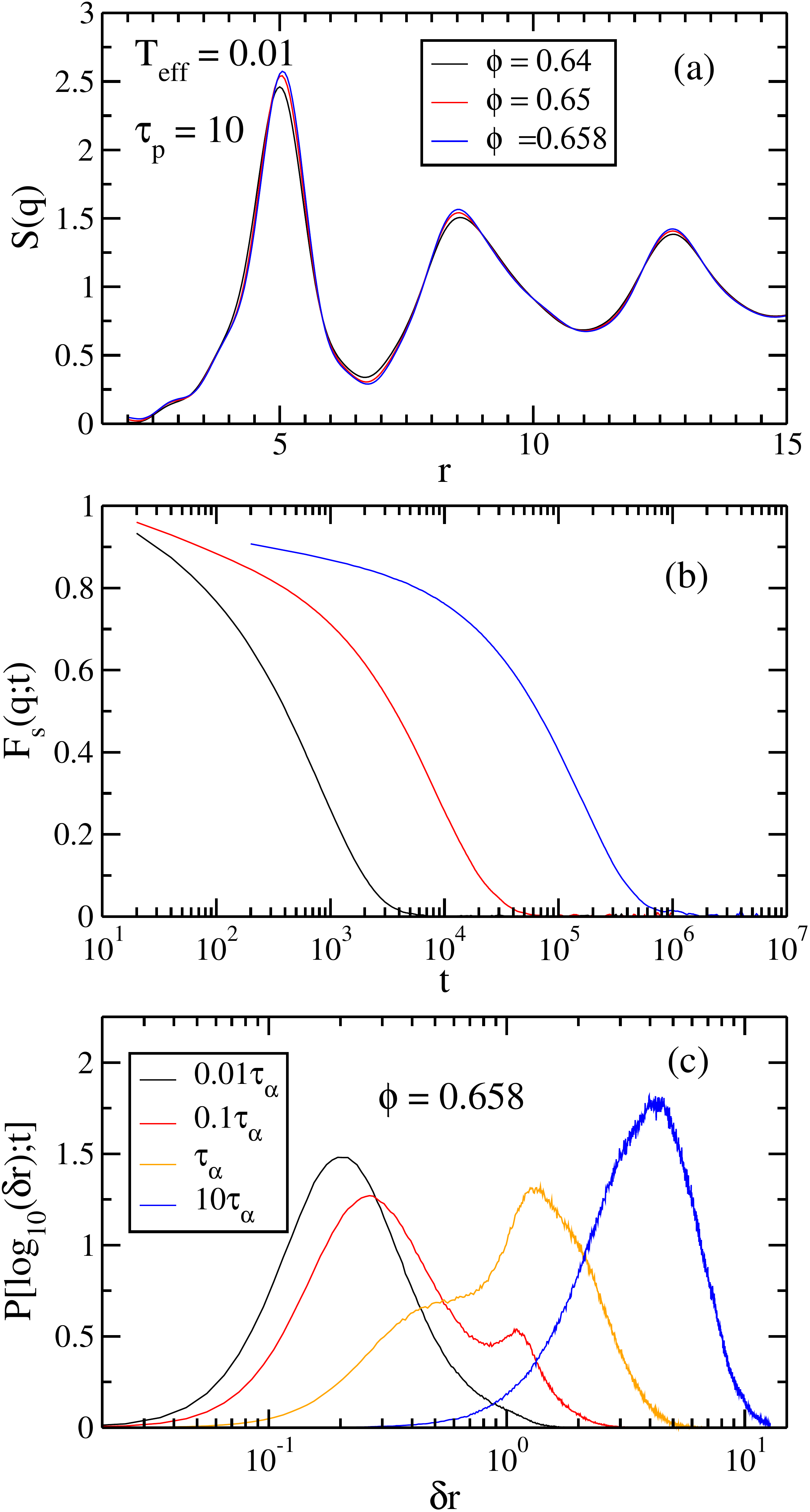}
\caption{\label{fig:noneq} 
Dense systems of self-propelled particles undergo a nonequilibrium glass transition as density is increased at constant effective temperature. Small variations of the steady state structure factor $S(q)$ shown in panel (a) are concurrent with dramatic slowing down of the relaxation of the self-intermediate scattering function $F_s(q,t)$ shown in panel (b), and the emergence of strong dynamic heterogeneity, which is evident form the probability distributions of the logarithm of the single-particle displacements, $P[\log_{10}(\delta r);t]$ shown in panel (c).}
\end{figure}

To elucidate the role of the activity we investigated the structure and dynamics of systems of AOUPs
with the WCA interaction. Since there are three control parameters, we fixed the effective temperature
at two illustrative values and then investigated the density and persistence time dependence of 
the structure and dynamics at each $T_{\mathrm{eff}}$. The two values of the effective temperature
correspond to two limits of the WCA interaction. At the higher temperature, $T_{\mathrm{eff}}=1.0$, the 
particles are able to explore a significant range of the repulsive part of the pair interaction. At the lower temperature, $T_{\mathrm{eff}}=0.01$, 
the particles do not penetrate the repulsive wall of the potential and they should behave effectively almost like hard
spheres. Thus, with these two values of $T_{\mathrm{eff}}$ we hope to analyze the behavior of a broad class of representative model systems, from models for dense liquids to dense assemblies of repulsive colloids and grains.

The central outcome of most numerical studies of dense systems with self-propulsion is that, as the strength of the self-propulsion is decreased, \textit{i.e.} as the effective temperature is decreased, or as the `crowding', \textit{i.e.} density is increased, the material undergoes a form of dynamic arrest characterized by a phenomenology very similar to observations in equilibrium systems driven by thermal fluctuations. We demonstrate these central observations in Fig.~\ref{fig:noneq} where we show the modest evolution of the pair structure of the AOUP model, which accompanies the dramatic slowing down of the dynamics and clear dynamic heterogeneity. In fact, to an unexpert eye, the data in Fig.~\ref{fig:noneq} could very well be taken as classic signatures of the glassy dynamics usually observed in equilibrium liquids, but they are reported here for a driven active system of self-propelled particles. 

\subsubsection{Nonequilibrium structure of active fluid}

Let us now turn to a more detailed description of the physics associated to nonequilibrium glassy dynamics of active particle systems. We argued in the previous paragraph that active materials display all classic features of equilibrium glassy dynamics. Thus, our goal here will be to emphasize the new features and difficulties that are specific to active systems.  

\begin{figure}
\includegraphics[scale=0.3]{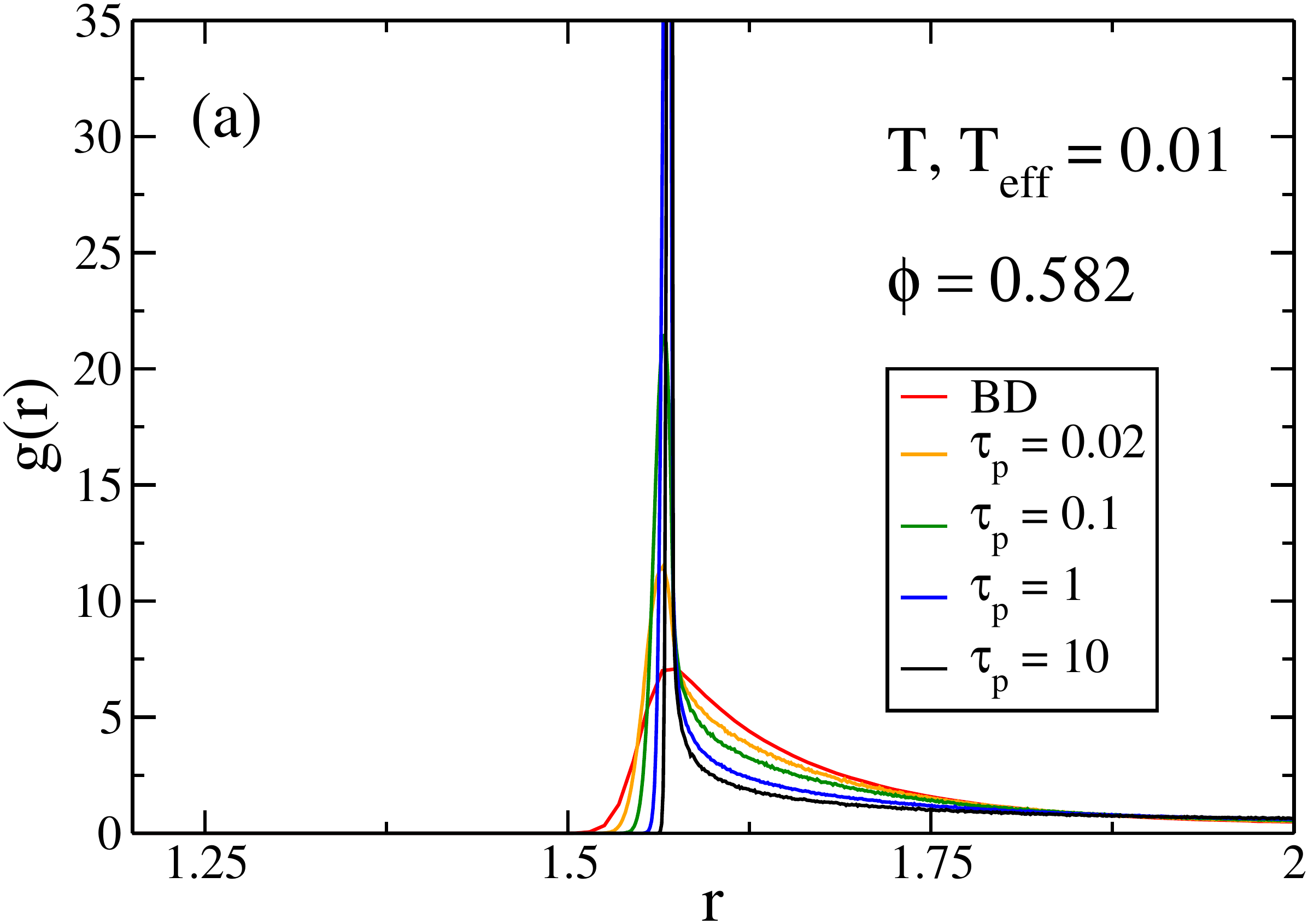}\\
\includegraphics[scale=0.3]{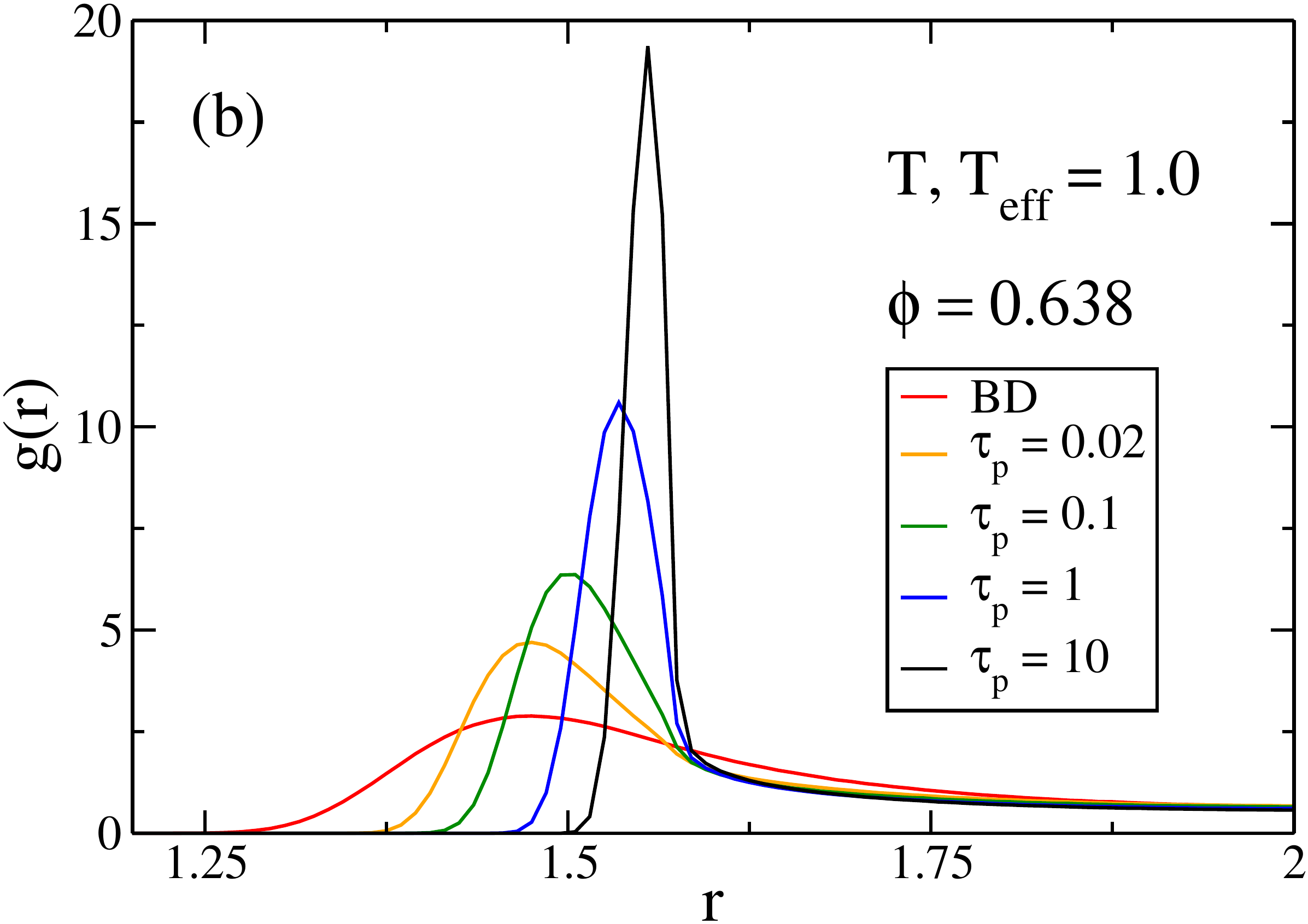}
\caption{\label{fig:agr} The steady-state pair correlation function $g(r)$ for an AOUP system with WCA interactions 
at (a) low effective temperature, $T_\mathrm{eff}=0.01$, and 
(b) high effective temperature, $T_\mathrm{eff}=1.0$. 
In the former case, particles almost never overlap but the self-propulsion leads to an effective attractive interation that makes the particles ``sticky''. In the latter case there is some interpenetration and some effective attractive interaction but the main consequence of the self-propulsion is the increase of the effective particle radius with $\tau_p$.} 
\end{figure}

We start with a description of the structure of the active system approaching the glass transition. For a fixed value of the persistence time, we have shown that the pair structure evolves very little as the glass transition is approached. It is, however, interesting to ask; how does the pair structure evolve as the system increasingly departs from equilibrium with increasing the persistence time?
In Fig.~\ref{fig:pstatics} we fix the value of the effective temperature, and show how the pair correlation function $g(r)$ changes as $\tau_p$ increases. Recall that as $\tau_p \to 0$, the system is at thermal equilibrium at a temperature $T_{\rm eff}$. 

We observe that the increasing activity has a profound
influence on the pair structure of nearest neighbors. The first peak of the pair correlation
function increases very rapidly with increasing persistence time, it reaches very large values and becomes extremely narrow at the lower effective temperature. We believe that equilibrium 
systems with similar short distance structure would be totally arrested. 

More in detail, we observe that at low $T_{\rm eff}$, the position of the first peak of $g(r)$ remains at the same distance $r$ corresponding to the cutoff of the WCA potential, and thus to the radius of the equivalent hard sphere system. The growing peak amplitude can be interpreted as an effective short-range attraction resulting from the competition between the repulsive interaction and the self-propulsion. This effective adhesion has been discussed in the context of motility-induced phase separation and cluster formation in self-propelled particles. 

For the system at higher $T_{\rm eff}$, the growth of the peak amplitude is observed but is less pronounced than for the hard sphere limit. This reflects a more subtle change in the effective interaction between particles. Perhaps the more striking observation is that the peak position is highly sensitive to the persistence time and shifts to larger distances as $\tau_p$ increases. Physically, this means that the effective radius of the particles is actually increasing as the persistence time grows, suggesting an increasing crowding of the particles. 

We will discuss below the dynamical phenomena that complement these observations. However, it is important to realize that the non-equilibrium nature of the self-propulsion dynamics implies that the sole knowledge of the interaction potential between particles is not enough to predict the structure of the non-equilibrium fluid. For equilibrium systems, a very accurate liquid state theory was developed decades ago to predict the fluid structure starting from the pair interaction~\cite{HansenMcDonald}. There exists at present no such theory for active matter, but we clearly observe that such theory should take into account the details of the self-propulsion mechanism.    

\subsubsection{A purely nonequilibrium object: velocity correlations} 

For equilibrium systems, the static structure is characterized by either $g(r)$ or by $S(q)$, which are the most
important structural quantities. In fact, almost all theories of glassy dynamics use the pair structure
as the only static input. 

An important development originating from the theoretical description of
dense active systems is the discovery that an additional correlation function appears in active systems that has no equilibrium analog~\cite{Szamel2015}. 
This function quantifies correlations of the 
velocities of the individual particles. The velocity of overdamped AOUP $i$ is equal to
$\xi_0^{-1}\left(\mathbf{f}_i+\mathbf{F}_i\right)$, recall Eq.~(\ref{eom1}), and the velocity correlation function
in Fourier space is defined as 
\begin{equation}\label{omegadef}
\omega_{||}(q) = \hat{\mathbf{q}} \cdot 
\left< \sum_{i,j=1}^{N}\left(\mathbf{f}_i+\mathbf{F}_i\right)\left(\mathbf{f}_j+\mathbf{F}_j\right)
e^{-i\mathbf{q}(\mathbf{r}_i-\mathbf{r}_j)}\right>\cdot\hat{\mathbf{q}},
\end{equation} 
with $\hat{\mathbf{q}} = \mathbf{q} / | \mathbf{q} |$. We note that for a binary mixture
there would be three different partial correlation functions of the overdamped
velocities. 

\begin{figure}
\includegraphics[scale=0.3]{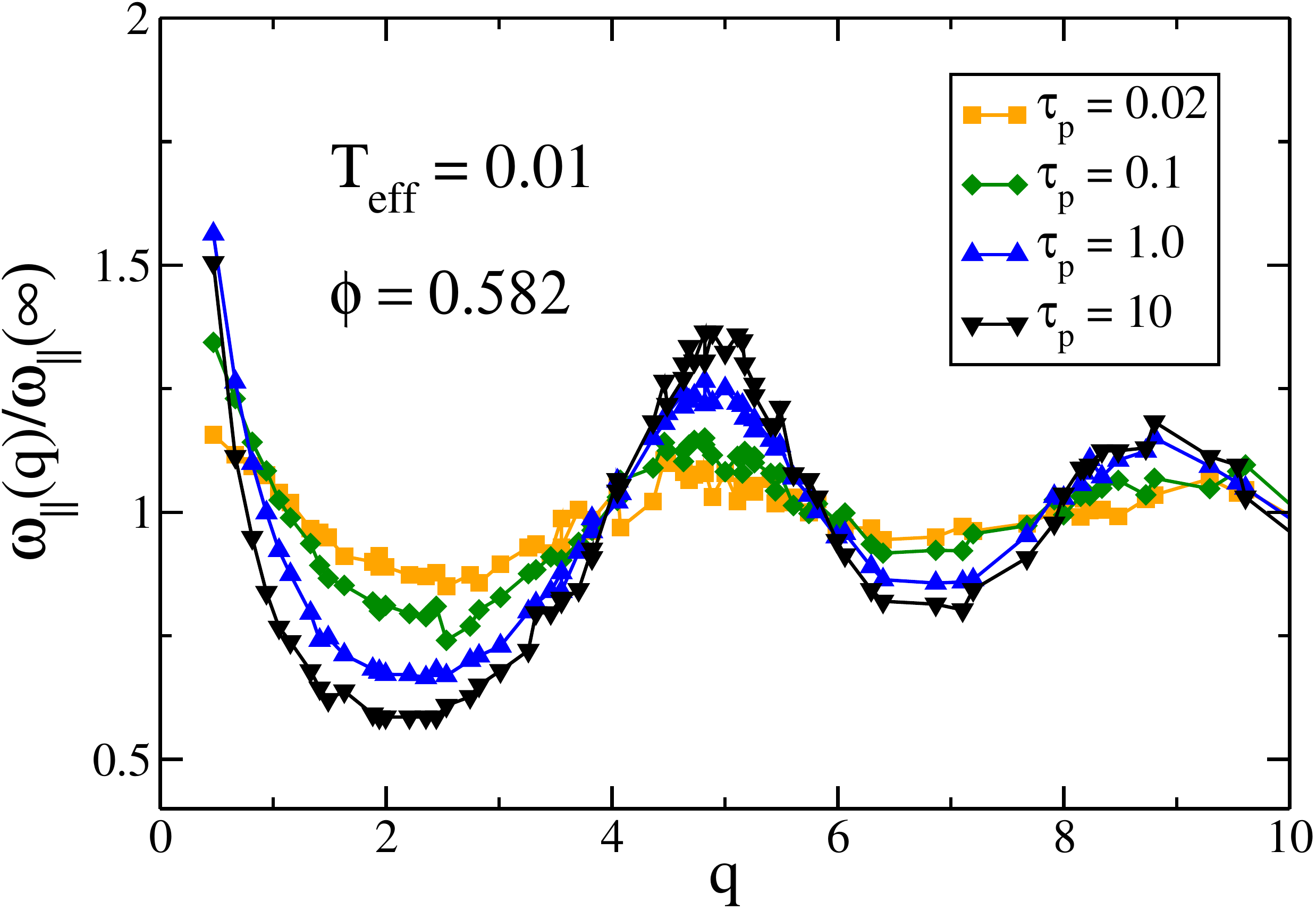}
\caption{\label{fig:aomega} The steady-state equal-time velocity correlations 
$\omega_\parallel(q)$ for an AOUP system with WCA interactions 
at a low effective temperature, $T_\mathrm{eff}=0.01$. Increasing range of the velocity correlations with increasing $\tau_p$ is signaled by the growth of the small wavevector peak of $\omega_\parallel(q)$. Developing local structure of the velocity correlations is evident from the growth of the amplitude of the oscillations of $\omega_\parallel(q)$.}
\end{figure}

In the limit of vanishing persistence time the correlation function in  Eq.~(\ref{omegadef}) becomes
trivial, \textit{i.e.} wavevector independent, because positions and velocities are independent quantities at thermal equilibrium. 
For finite persistence times it has a non trivial wavevector
dependence, as shown in Fig.~\ref{fig:aomega}. In addition to the large $q$ oscillations that imply local velocity correlations reflecting the local structure of the dense liquid, there is a clear upturn at low-$q$ that can be used to define a finite correlation length for velocity correlations.  

Physically, the non-trivial character of the velocity correlation function implies that a snapshot of short-time displacement fields is likely to reveal large-scale correlations that are purely due to the non-equilibrium nature of the active particle system. These spatial correlations represent a non-trivial form of collective motion. We note that these correlations exist even in the dense, but non-glassy, active liquid and are thus not specifically connected to the glassy dynamics itself. Numerical measurements indicate that the temperature dependence of the velocity correlation function is relatively modest, suggesting that correlations already present in the active fluid survive but do not change in any remarkable way as the non-equilibrium glass transition approaches. 

The theoretical importance of the velocity correlations (\ref{omegadef}) is twofold. First, these correlation functions enter into the exact description of the short time dependence of various
correlation functions. Second, they also enter into approximate theories of the long-time dynamics of active glassy systems. 

\subsubsection{How does activity change the slow dynamics?}

We now turn to the dynamics. We note that, quite surprisingly, in some cases the evolution of the relaxation time for a fixed $T_{\mathrm{eff}}$ does not change monotonically with $\tau_p$~\cite{Szamel2015}. For small $\tau_p$ 
the relaxation time initially decreases and then increases
with increasing $\tau_p$. This finding demonstrates that the activity can alter
the glassy dynamics in rather subtle, unexpected ways. This non-monotonic 
behavior of the relaxation time is not mirrored in structural 
quantities such as $g(r)$ and $S(q)$, since for instance  
the height of the first peak of $g(r)$ increases monotonically with persistence
time, even though the relaxation time does not. 

An enhancement of the structure, as given by the increase of the peak height of $g(r)$ and a decrease of the relaxation time is contrary to what is expected from studies of equilibrium glassy liquids. A well-studied theory for passive liquids that connects the 
liquid structure with dynamics is the mode-coupling theory~\cite{Goetzebook}, which has only the
static structure factor as input. While the mode-coupling theory for the glass
transition is not an exact description for the glass transition, it can make  predictions and it provides microscopic physical insights.  

To gain some insight into why active systems may have a non-monotonic evolution
of the relaxation time with the persistence time, a mode-coupling-like theory 
for active systems was developed~\cite{Szamel2015,Szamel2016}. It was shown that if the theory incorporated the nontrivial character of the velocity correlations, the theory could indeed predict a non-monotonic evolution of the relaxation time with increased persistence time. There is a minimum of the relaxation time with increasing persistence time and the relaxation time begins to increase
again with increasing persistence time. The additional velocity correlations are, therefore, an important component of the slow dynamics of dense active systems.

\begin{figure}
\includegraphics[scale=0.3]{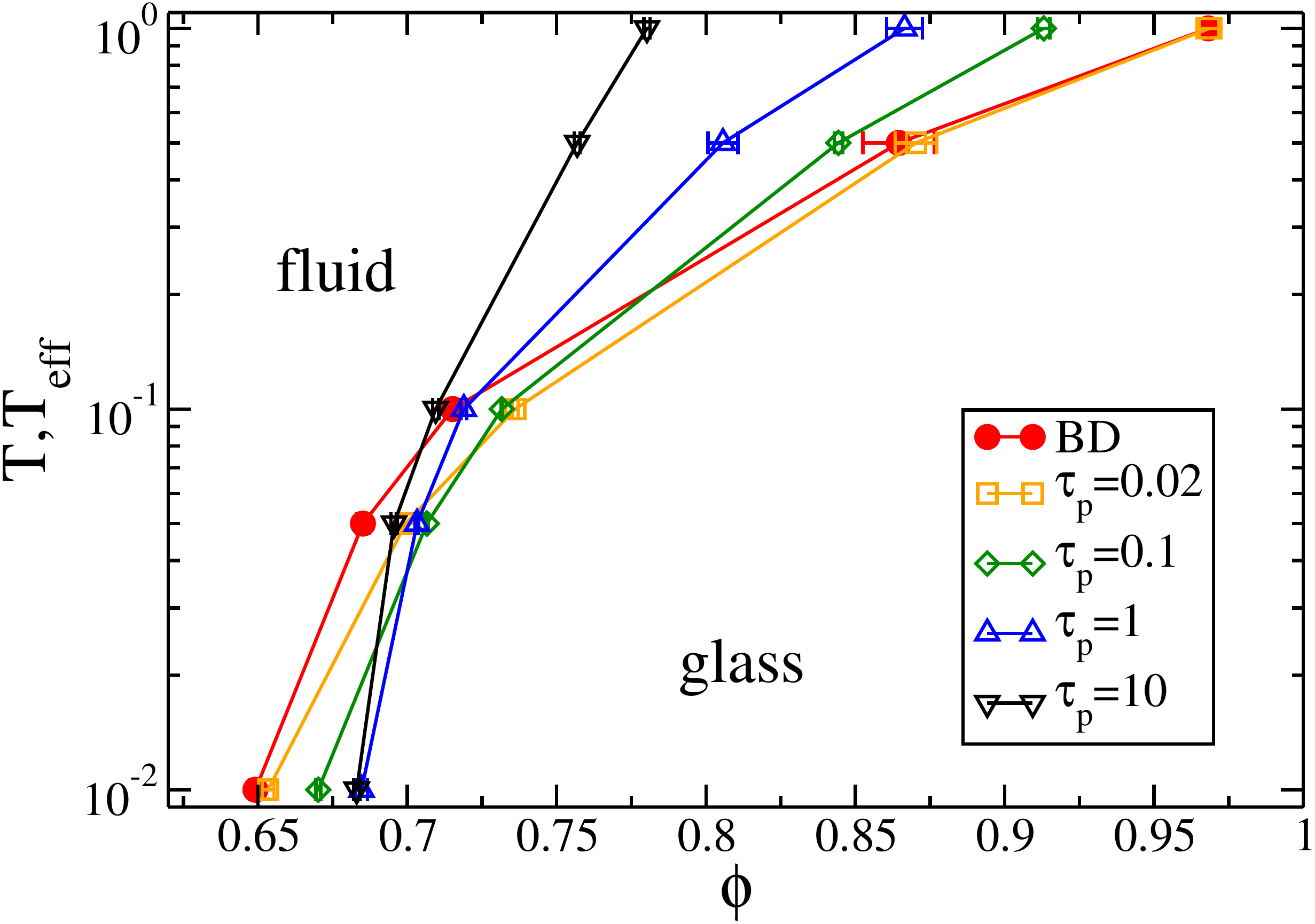}
\caption{\label{fig:glasspd} Evolution of the fluid-glass phase diagram with the persistence time of the self-propulsion.  With increasing persistence time the glass transition line shifts towards smaller volume fractions at higher effective temperatures and towards larger volume
fractions at low effective temperatures, so departure from equilibrium can either promote or suppress the glassy dynamics. Filled symbols are Brownian dynamics (BD) simulations.} 
\end{figure}

Next, we focus on the glass transition itself~\cite{Flenner2016,glassline}. For a fixed value of the persistence time, we find that the relaxation time increases when the effective temperature is decreased and/or when the density increases, just as for dense equilibrium fluids. This simply means that even for active systems, the glass transition results from a competition between active forces that make the particles move, and crowding that tends to arrest them. 

From the increase of the relaxation time of the system, we use an empirical fitting form, $\tau_\alpha \sim \tau_0 \exp (B / (\phi_0 - \phi))$ to obtain a critical density for the glass transition, $\phi_0(T_{\rm eff}, \tau_p)$, which depends on the other two parameters of the model. The evolution of the glass transition lines are reported in Fig.~\ref{fig:glasspd}, in a (temperature, density) phase diagram. For a given value of the persistence time, the phase diagram offers two phases, the fluid at low density and high temperature, and the glass at large density and low temperature. The `BD' line is obtained from simulations performed with Brownian dynamics, i.e. in the equilibrium $\tau_p \to 0$ limit, and it corresponds to the equilibrium glass transition. All other lines correspond to non-equilibrium glass transition lines. 

The influence of a finite persistence time on the glass transition is obvious. These data confirm that at low $T_{\rm eff}$ an increase of the persistence time shifts the glass transition towards large densities, whereas the opposite effect is observed for larger $T_{\rm eff}$, with a complex behavior at intermediate $T_{\rm eff}$ values. These non-trivial dependencies show that departing from equilibrium can either promote or depress glassy dynamics, and that it is difficult to form a physical intuition, even for very simplistic models such as AOUPs. 

\section{More complex models}

In this section we briefly discuss some of the more complex computational models of dense active matter.
Most of these models were proposed to incorporate specific features of activity encountered in laboratory
active matter systems. Some of them attempt to (semi-)qualitatively model specific biological, colloidal and
granular systems.

\subsection{More models for self-propelled particles}

Perhaps the most popular model for active particulate systems is that of active Brownian particles~\cite{tenHagen,FilyMarchetti}. 
It models active matter as consisting of particles that move via a combination of active, directed motion
and Brownian motion. The particles are endowed with an axis of symmetry. They move systematically along
this axis with a constant velocity $v_0$. The direction of the axis of symmetry moves via rotational diffusion with diffusion coefficient $D_R$. In addition, the particles are also subjected to a random
Brownian force and instantaneous friction characterized by temperature $T$ and friction coefficient $\xi_0$. The particles interact via spherically symmetric interaction potential $V(r)$. 
In the original version of the model, the resulting
translational diffusion coefficient due to these random forces, $D_T=T/\xi_0$ and the rotational diffusion coefficient were constrained to follow 
the relation imposed by hydrodynamic considerations~\cite{tenHagen}, 
$D_R= 3D_T/\sigma^2$, 
where $\sigma$ is the particle diameter. In several studies this relation has been relaxed~\cite{Ni} and
both $D_R$ and $D_T$ were treated as independent parameters, which is equivalent to using the persistence time of the AOUP particle  as a free parameter of the model.

The active Brownian particles 
model is intended to represent active colloids. It has been used in particular in many studies of
motility-induced phase separation. It has also been used to study the influence of the activity on
the glassy dynamics. In a simulational study of active Brownian hard spheres, 
Ni \textit{et al.} \cite{Ni} showed that when the magnitude of the systematic velocity is increased while all the other parameters are kept constant, the apparent glass transition volume fraction moves towards larger values. 
Ni \textit{et al.} noted that the faster dynamics was accompanied by decreasing height of the first peak
of the steady state structure factor. This finding qualitatively agrees with the results of the investigation of the glassy phase diagram described in the previous section, where this corresponds to low $T_{\rm eff}$ values for the AOUP model. In another study, Fily {\it et al.}~\cite{Fily2014} 
used the active Brownian particle model with a soft repulsive potential to map out the density-temperature phase diagram of the model. They also reported a `frozen' phase at low activity and large density, which in our view should be interpreted as a glass, but the slow glassy dynamics on the approach to this arrested glass phase was not analyzed in detail. We expect that it should present the same phenomenology as the AOUP models shown in Sec.~\ref{minimal} above.

\subsection{Aligning interactions}

The field of active matter was largely born from the quest to describe and understand theoretically the physics of animal flocks. The Vicsek model \cite{Vicsek} was conceived to capture the competition between the natural tendency for animals to align the direction of their self-propulsion and an external noise. While it is unclear whether all self-propelled particles types (such as cells and colloidal particles) truly possess the same tendency to alignment, the existence of \textit{implicit} aligning interactions was demonstrated for some active materials, such as vibrated polar disks \cite{Deseigne}.

A computational model was proposed in Refs.~\cite{model-olivier,model-olivier2} to 
describe a system of vibrated polar disks first studied in Ref. \cite{Deseigne}. 
In this computational model, the motion of the particles is not overdamped. The particles are endowed with a polarity represented by a unit vector. Thus, the instantaneous state of a given particle is represented by its 
position, velocity and polarity. The self-propelling force of constant magnitude acts in the direction
of the polarity. In turn, there is a torque acting on the velocity, which tends to align it with the polarity.
There might also be two stochastic torques that randomly rotate the velocity and the polarity vectors.
Finally, the particles interact via a spherically symmetric interaction potential. This is quite a complicated model with many adjustable parameters that leads to a huge parameter space. However, since it was first
proposed to describe features observed in a specific experimental work, 
the parameters where adjusted to best reproduce the results of that specific experiment. It was shown that single particle, binary and collective properties of the experimental system can indeed be reproduced numerically.
Both the experimental and the model have also then been studied at larger density when particles form an active crystalline phase~\cite{briand}. 
Computational and experimental studies of the glassy phase of a binary mixture of the same model are currently in progress and preliminary results suggest that an active glassy phase is indeed found, whose properties will hopefully be analyzed in more detail in future work. We note that the model of Refs.~\cite{model-olivier,model-olivier2} could be thought of as an under-damped version of a model analyzed in Ref.~\cite{Henkes2011}. The latter model also exhibits implicit aligning interactions. The authors of Ref.~\cite{Henkes2011} identified a `jammed' phase that in our view is an arrested glass phase. Again the transition between the fluid and arrested phases was not characterized in any detail, and this would be a worthwhile research effort.  

In an effort to describe the collective motion observed in dense epithelial tissues, Sepulveda {\it et al.}~\cite{hakim} proposed a computational model where particles interacting with a rather complex pairwise interaction are self-propelled with a finite persistence time and are subject to short-range aligning interactions between the direction of the self-propulsion. Again, the parameter space of the model is impressive, but the many parameters of the model were adjusted to reproduce a specific set of experimental observations. In some later versions of the model, friction to a substrate and additional ingredients were added to the model~\cite{Garcia2015}. Finally, we note yet another model with aligning interactions introduced in Ref.~\cite{cerbino}.
It would be interesting to try and simplify such models in order to address more specifically the physical question of how aligning interactions between self-propulsion directions may affect, and perhaps change qualitatively the glassy dynamics obtained in the absence of aligning interactions reviewed above in Sec.~\ref{minimal}. 

\subsection{Modeling cell dynamics: division, death, and volume fluctuations}

One of the common phenomena in biological active matter is cell division and death. A combination of these
processes can lead to an unstable system with cells eventually dying out, or instead a growing tissue that expands and invades space. If cell death and division are instead statistically balanced, a driven steady state can be reached. Quite importantly, for the present article, it was shown by Matoz-Fernandez \textit{et al.}~\cite{Matoz} that cell division and death can strongly influence the glassy behavior. 
In this study, a particle-based model of two dimensional ephitelial tissues was investigated. 
The particles interacted via a combination of a short-range repulsive and a longer-range attractive harmonic 
potentials. The activity was modeled as a combination of a cell death process, in which particles representing
cells were randomly removed from the system and a cell division process, in which a new (daughter) cell
was added on top of an existing (mother) cell, with a probability depending on the number of neighboring
cells in contact with the mother cell. 

A rich non-equilibrium phase diagram with gas-like, gel-like and dense 
confluent phases was found. A remarkable result is that in the dense, confluent phase any positive rate of cell death and division always fluidizes the system and prevents any amorphous solidification. Physically the reason is that any such event reorganizes the system locally, and thus at long times any location in the system has eventually reorganized with probability unity, completely reshuffling the structure; the dynamics is not arrested. 
In contrast, a system without cell death and division but endowed with motility-like activity modeled
similarly to that present in the active Brownian particles model, was found to exhibit classic features of
glassy dynamics upon decreasing the magnitude of the velocity, as expected by analogy with the type of minimal active model discussed in Sec.~\ref{minimal}.


In the opposite limit where death does not compensate by cell division the density of cells would increase exponentially with time in a confined volume.  Here the interesting setting is when open boundary conditions are present, since from just a few cells that can divide a large tissue/colony can expand. 
This was numerically modeled in Ref.~\cite{Malmi} using an appropriate pairwise interaction and a dynamics uniquely ruled by the stochastic rules for particle division. In agreement with the steady state study of  Matoz-Fernandez \textit{et al.},  Malmi-Kakkada \textit{et al.} \cite{Malmi} conclude that cell division also leads to a complete reshuffling of the growing colony at long times, suggesting that cell division rate directly controls the onset of glassy dynamics. However, they surprisingly do not observe any specific glassy feature even when the division rate is small. It would be interesting to specifically revisit such a model in order to analyze in more detail the microscopic mechanisms responsible for tissue fluidisation~\cite{ranft,prost}.

Finally, a qualitatively different type of active dynamics, inspired by observations of real tissue dynamics, has recently been numerically studied~\cite{elsen}. In this model, the confluent tissue is again modeled as soft repulsive particles at large density, and the only source of activity is given by spontaneous fluctuations of the particle volume. Observations in dense epithelial tissues suggest that individual cells undergo relatively large volume fluctuations (up to 20\%) that appear almost periodic with a very low frequency~\cite{volume}. In the numerical model, these oscillations were taken as purely periodic with a very low but fixed frequency. In the opposite limit where frequencies are widely distributed or changing with time, the driving becomes random and resembles the random fluctuations provided by thermal noise, thus leading to ordinary equilibrium-like glassy dynamics. For purely periodic driving forces, a sharp fluid-to-solid transition is reported, but it shares no features with the glassy dynamics. Instead, there seems to exist a sharp threshold between an arrested state when volume fluctuations are small to a fully fluidized state when they are large. The transition between the fluid and solid states appears discontinuous, and is akin to a non-equilibrium first order transition. It was argued that the proper analogy with the physics of glasses is not with the glass transition itself, but rather to the yielding transition discussed in Sec.~\ref{sec:rheology} which is also found to be discontinuous~\cite{takeshi}. Physically, volume fluctuations appear as a slow driving force, and fluidization occurs when that force exceeds a threshold, as for yielding, the only difference being that the force acts on a local rather than a global scale. This transition thus qualifies as an `active yielding transition'.

\subsection{Vertex models for tissue morphology}

Finally, let us briefly discuss two models that, to different degrees, are not particle based.
These models belong to the category of vertex-like models that are very popular among researchers
focusing on modeling real confluent biological tissues. The first model is the so-called Voronoi 
model~\cite{biPRX}.
In this model, the cells are modeled as Voronoi volumes defined by their neighbors and the degrees of freedom
are the Voronoi cell centers. However, the energy expression is that of the standard vertex model~\cite{vertex}, 
where the energy is given as the sum of quadratic departures of the area and the perimeter from
their preferred values. In this model the forces act on the Voronoi cell
centers. The second model is the standard vertex model~\cite{vertex}, in which the degrees of freedom (on which the
forces and thermal noise are acting) are the positions of the vertices of each cell. The same energy expression 
is used as in the Voronoi model.  

Vertex models are interesting models because they reflect more faithfully the geometric structure of dense confluent tissues. A remarkable result is that the vertex model may undergo a jamming transition in the absence of driving that is purely controlled by the competition between surface and bulk terms in the energy functions~\cite{biNature}. Therefore, as the average shape of the cells evolves the mechanical response of the system changes from a fluid to a solid response, in very much the same way a dense packing of soft particles undergoes a jamming transition as the density is increased, as discussed in Sec.~\ref{sec:jamming}. 

The properties of these models have also been studied in the presence of either thermal forces (\textit{i.e.} in equilibrium), or in the presence of self-propulsion with a finite persistence time~\cite{biPRX,daniel,lisa}. For a given persistence time (that can be zero), a transition between a fluid and an arrested solid state is observed, with a growing relaxation time and, again, the phenomenology associated to a nonequilibrium glass transition suggesting that a phase diagram for vertex models in a plane comprising activity and parameter shape should qualitatively resemble the sketch in Fig.~\ref{fig:rhot} for soft spheres. Further work should clarify the details of both the glass and the jamming transition in the broad family of vertex and Voronoi models. 

\section{Conclusions and some theoretical perspectives}

Among the many directions that have emerged in the growing field of active matter, the analysis of dense systems composed of individual entities locally driven by active forces is receiving attention from a large community of scientists, and novel experiments and active materials in this regime keep emerging. In this work, we have provided a conceptual framework to understand how dense active materials may become dynamically arrested when active forces lose the battle against particle crowding to form an amorphous state of active matter.  We have argued that systems such as dense tissues, self-phoretic colloids and active granular materials would display similar glassy dynamics despite the fact that they evolve far from equilibrium. This observation suggests that the field of the equilibrium glassy dynamics and the field of dense active materials are intimately connected.  

We argued that minimal models of active particle systems confirm the experimental observation that a fluid to amorphous solid glass transition can indeed be observed, for instance using computer simulations. The simplicity of these models, as compared to the complexity of the experimental realizations described above, makes these models useful starting points to address very precise questions about the modifications brought by the non-equilibrium nature of active matter to the equilibrium glass transition phenomenon.   

There are many challenges ahead of us to get a better theoretical understanding of non-equilibrium glass transitions. 
The main conceptual difficulty is that 
active systems are out of equilibrium. This makes their analysis quite delicate as the tools of equilibrium statistical mechanics are no longer available. 
As a result, there exists at present no established theory to describe the steady state structure of simple active fluids, not to mention very dense active fluids. Furthermore, the
various identities relating equal-time and time-dependent correlation functions and static and dynamic
linear response functions become invalid, and there might be 
systematic currents inside the active system.

The theoretical description of dense active matter systems is in its infancy. There have been a few approaches
to describe the steady state properties of dense active matter 
systems~\cite{Maggi,Binder,FarageKB,Maggiveldis,Maggipressure,MaggiBrader1,MaggiBrader2}.
These works focused on the influence of the activity on the phase behavior. Reasonable theoretical descriptions
of the influence of the activity on the pair structure and of the phase behavior were obtained. However,
Rein and Speck~\cite{ReinSpeck} 
questioned this approach and the general ability of effective pair potentials to describe the
local structure. Speck, L\"owen and collaborators
developed their own approach to the phase behavior that more explicitly involved self-propulsion and correlations
involving the self-propulsions~\cite{BialkeLS}. 
They also generalized the dynamical mean-field theory to include the activity
and derived an active version of the Cahn-Hillard theory~\cite{SpeckCH1,SpeckCH2}. 
Finally, L\"owen and collaborators generalized dynamical density functional
theory to active particles~\cite{LowenADDFT1,LowenADDFT2,LowenADDFT3}.

Of particular interest to us are the theories for the active (nonequilibrium) glassy 
dynamics~\cite{BerthierKurchan,Farage,Szamel2015,Szamel2016,Feng2017,Voigtmann2017,NandiGov2017,NandiPNAS,Szamel2019}.
These theories are inspired by theories of equilibrium glassy dynamics. Specifically, there is a 
theory~\cite{BerthierKurchan} that follows the spirit of the mean-field $p$-spin approach \cite{pspin}, 
a number of theories~\cite{Farage,Szamel2015,Szamel2016,Feng2017,Voigtmann2017,NandiGov2017,Szamel2019} that 
rely upon a factorization approximation that is the cornerstone of the mode-coupling theory of glassy
dynamics~\cite{Goetzebook}, and a theory~\cite{NandiPNAS} that attempts to include some elements of active matter into the framework of mean-field (RFOT) glass theory~\cite{RFOT}. 


Theories in this second group adopt a more microscopic perspective, \textit{i.e.} they deal with 
systems of moving particles and include at least some static (equal-time) correlations. These theories 
systematically predict the existence of a non-equilibrium form of a glass transition. Although we argued that this 
is a generic result, we recall that the same theories would predict that the glass transition of mechanically 
driven glasses would be destroyed by the external mechanical constraint, suggesting that the effect of active or mechanical driving on glasses can be rather subtle. 

The theories in the second group can be differentiated using several criteria. First, these theories can be 
differentiated with respect to the model they attempt to describe. Thus, there is a 
theory developed by us that focuses on the simplest model system, the system of athermal 
AOUPs~\cite{Szamel2015,Szamel2016}, a theory that describes a thermal AOUP system (\textit{i.e.} 
a system of particles subject to both self-propulsions evolving according to the Ornstein-Uhlenbeck process and
random thermal forces) and three theories~\cite{Farage,Voigtmann2017,Szamel2019} describing active Brownian particles. 
In our opinion, it is relatively easy to adopt a theory developed for one model system to another model system
and the more substantial differences between various theories of mode-coupling flavor originate from another source. 
Second, we can contrast between theories that 
attempt to describe both steady-state properties and the dynamics of active particles 
\cite{Farage,Voigtmann2017} using a version of the ``integration-through-transients'' approach 
used to describe the properties of sheared colloidal suspensions~\cite{ITT}  
and theories that assume that the steady-state structure is known and that focus instead directly on the prediction of stationary time correlation functions~\cite{Szamel2015,Szamel2016,Feng2017,Szamel2019}. Third, we can contrast 
theories that attempt to replace the active model system by an equivalent equilibrium (thermal, passive) 
system~\cite{Farage,Feng2017} and theories in which the time-evolution of the self-propulsion is accounted for 
explicitly albeit approximately~\cite{Szamel2015,Szamel2016,Voigtmann2017,Szamel2019}. 

It seems that important ingredients of a theory are a reasonably accurate description of
the time evolution of the self-propulsion and incorporation of new nonequilibrium static correlations. 
Thus a promising route for future work seems to be a combination of the formally exact description of the time evolution
of the self-propulsion of the approach of Ref.~\cite{Voigtmann2017} and of our own theory~\cite{Szamel2015,Szamel2016} which automatically takes into account nonequilibrium static correlations.

We have shown that a nonequilibrium glass transition is a general feature of model active matter
systems. In the future, on the experimental and simulational side, the focus should shift to investigating specific 
nonequilibrium features of the transition. For example, since it was shown that polar driven grains can form
a flowing crystal~\cite{BriandSchindlerDauchot}, it would be interesting to investigate whether and under what
conditions a flowing
glass state is possible. On the theoretical side, the validity and accuracy of various theories should be 
quantitatively assessed. 
To this end one needs to carefully simulate model active systems, measure various static correlations
and then use them in mode-coupling equations in order to determine the approximate dynamics. The results should
then be compared with the simulations. This would parallel the work done previously 
in the area of equilibrium glassy dynamics~\cite{Flenner2005b,Weysser}.

\section*{Acknowledgments}

GS and EF gratefully acknowledge the support of NSF Grant No.~CHE 1800282.
The research leading to these results has received funding from the Simons Foundation (\#454933, L. Berthier).

\end{document}